\documentstyle[11pt,newpasp,twoside,psfig]{article}
\def\edcomment#1{\iffalse\marginpar{\raggedright\sl#1\/}\else\relax\fi}
\reversemarginpar

\def\sn{SN\,1987A}
\def\kms{km~s$^{-1}$}
\def\HII{H\,{\sc ii}}

\begin{document}
\title{Supernova 1987A: A Young Supernova Remnant in an Aspherical
Progenitor Wind}
 \author{B. M. Gaensler\altaffilmark{1}}
\affil{Center for Space Research, Massachusetts Institute of Technology, \\
70 Vassar Street, Cambridge, MA 02139, USA}
\author{R. N. Manchester, L. Staveley-Smith, V. Wheaton,
A.~K.~Tzioumis, 
J.~E.~Reynolds, M. J. Kesteven}
\affil{Australia Telescope National Facility, CSIRO, PO Box 76,
Epping, NSW 1710, Australia}
\altaffiltext{1}{Hubble Fellow}

\begin{abstract}
The interaction between the ejecta from Supernova 1987A and surrounding
material is producing steadily brightening radio and X-ray emission.
The new-born supernova remnant has been significantly decelerated by
this interaction, while its morphology reflects the axisymmetric nature
of the progenitor wind.
\end{abstract}

\section{Introduction}

The circumstellar material around \sn\ shows striking deviations
from spherical symmetry, in particular in the form of the
``three-ring circus'' spectacularly imaged by {\em HST}\ (Burrows et al.
1995). This nebulosity shows a distinct bipolar structure, resembling
many of the planetary nebulae shown at this meeting. In the case
of \sn, supernova ejecta are rapidly propagating outwards
from the center of this structure, producing radio, optical and X-ray
emission as they collide with surrounding material. Observations
of \sn\ are thus an excellent probe of the mass-loss history of a supernova
progenitor.

Many authors have considered the nature of the triple-ring system
surrounding \sn. For the purposes of interpreting
the interaction between the supernova ejecta and this material,
we adopt the ``standard model'' (Blondin \& Lundqvist 1993; Martin
\& Arnett 1995), namely that:

\begin{itemize}
\item the progenitor star was a red supergiant (RSG) until $\sim$20\,000~yr,
during which time it produced a slowly moving, dense wind;
\item the star then evolved into a blue supergiant (BSG), producing
a fast moving and low density wind;
\item the rings correspond to the bipolar interface produced
by the interaction between the RSG and BSG winds;
\item the RSG wind was densest in the equatorial plane (perhaps
produced by rotation and/or binarity in the progenitor), while
the BSG wind was isotropic.
\end{itemize}

We note that this model certainly has its problems, and that many
alternatives have been proposed (e.g.\ Soker 1999).

\section{Radio Flux Monitoring}

Radio emission was detected from \sn\ just 2 days after the supernova
explosion (Turtle et al. 1987). This emission peaked on day 4, before
following a power law decay to become undetectable by day 150 (Ball et
al.  1995). This radio outburst as been interpreted as synchrotron
emission produced as the supernova shock passed through the innermost
regions of the BSG wind (Storey \& Manchester 1987).

After $>$3 years of radio silence, radio emission was re-detected from
\sn\ in mid-1990 (Staveley-Smith et al. 1992).  Since then, emission
has shown a monotonic increase with a spectral index $\alpha \approx -1$
($S_\nu \propto \nu^{\alpha}$) (Gaensler et al. 1997). X-ray emission
from the system turned on at around the same time, and has since also steadily
increased (Hasinger, Aschenbach \& Tr\"{u}mper 1996).  This behavior
suggests that the shock, having freely expanding through the
BSG wind, has now run into a density jump.

\section{Radio Imaging}

Observations of \sn\ with the Australia Telescope Compact Array
(ATCA) at 9~GHz can resolve the 
radio emission from \sn. By fitting a thin spherical
shell to the radio emission at each epoch, the expansion of the source
with time can be quantified. These data, shown in Figure~1,
show a linear expansion rate of $\sim$3500~\kms\ from day 1500 onwards.
Interpolating between day 0 and day 1500 implies an initial expansion
rate $>$35\,000~\kms, consistent with VLBI and H$\alpha$ measurements
made shortly after the explosion. These data thus indicate that the
supernova shock experienced a rapid deceleration at or just before
radio and X-ray emission were redetected in mid-1990.

\begin{figure}[ht]
\centerline{\psfig{figure=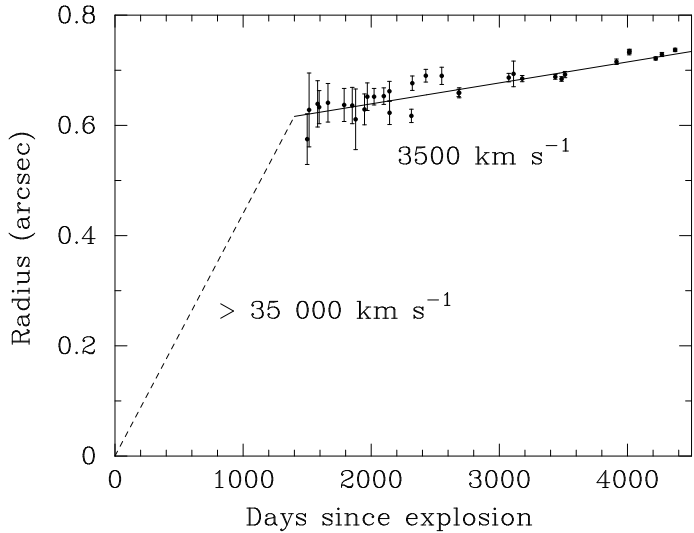,height=7cm,angle=270}}
\caption{The radius of the radio remnant as a function of time.}
\end{figure}

The diffraction-limited resolution of the ATCA is $0\farcs9$, but
using super-resolution we can produce a sequence of radio images
of \sn\ with a slightly higher resolution of $0\farcs5$ (see Gaensler
et al. 1997). These images (see Figure~2) show the emission to
have a shell-like structure; the morphology is dominated by two lobes
to the east and west.

\begin{figure}[ht]
\centerline{\psfig{figure=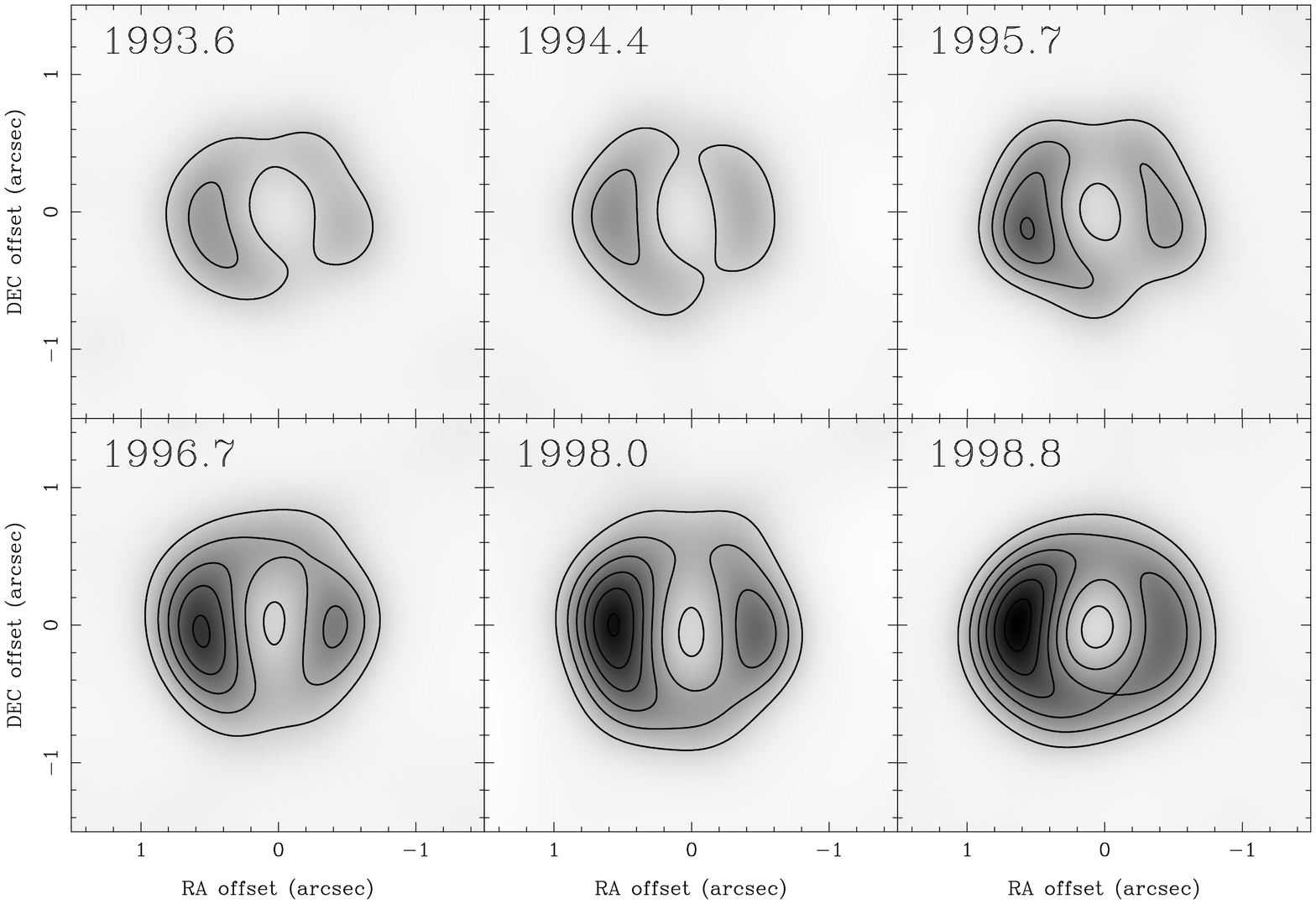,height=7.5cm,angle=270}}
\caption{Super-resolved ATCA images of SN\,1987A.
Contour levels are from 0.5 to 3.0~mJy~beam$^{-1}$, at
0.4~mJy~beam$^{-1}$ intervals.}
\end{figure}

An overlay between the optical ring system and the radio data
(Reynolds et al. 1995; Gaensler et al. 1997) shows the radio shell to
be centered on the position of the supernova, but with a radius only
$\sim$90\% of that of the optical ring.  Thus although the supernova
shock appears to have run into a density jump, this jump must be
located within the interface between the RSG and BSG winds.  The
radio/optical overlay also shows that the radio lobes align with the
major axis of the optical rings, when projected onto the sky.  Gaensler
et al. (1997) interpret this as indicating that radio emission is
confined to the equatorial plane of the progenitor system, a result
recently confirmed by STIS data (Michael et al. 1998).

\section{Interpretation}

The abrupt radio and X-ray turn-on in mid-1990, as well as the rapid
deceleration of the shock at around the same time, can be explained in
terms of the ``standard'' interacting winds model, with the addition of
a dense \HII\ region just inside the bipolar interface (Chevalier \&
Dwarkadas 1995).  This region, produced by UV photons from the
BSG ionizing the
surrounding RSG wind, can, at least to first order, account for the 
observed light
curves, expansion rate and X-ray emission measure.
The double-lobed morphology is then interpreted as an axisymmetry
in this surrounding material, as discussed in detail by Gaensler
et al. (1997).

\section{The Future}

Radio monitoring and imaging of \sn\ will certainly continue;
in 2001 the ATCA will be upgraded to
a maximum frequency of 100~GHz, giving a significant improvement
in spatial resolution. Meanwhile, {\em Chandra}\ and {\em XMM}\
will soon spatially and spectrally resolve the X-ray emission
from \sn, giving us a wealth of information about the conditions
at the shock. All this is just a prelude to the collision
of the supernova ejecta with the inner optical ring, expected
in around 2004. At this point \sn\ will drastically evolve
and brighten (perhaps
by a factor of $10^3$ in every waveband!),
providing us with much new information about the progenitor's
circumstellar material. Into the next century and beyond, we can
expect that \sn\
will evolve into a ``classical'' supernova remnant (SNR), at which point
we can perhaps start to relate the complex morphologies of SNRs
to the mass-loss histories of their progenitors.

\acknowledgments

The Australia Telescope is funded by the Commonwealth of Australia for
operation as a National Facility managed by CSIRO.  B.M.G.
acknowledges the support of NASA through Hubble Fellowship grant
HF-01107.01-98A awarded by the Space Telescope Science Institute.

\end{document}